\documentclass[aps,pra,graphicx,9pt,onecolumn,superscriptaddress,twocolumn]{revtex4-2}
\usepackage[colorlinks,linkcolor=blue,urlcolor=blue,anchorcolor=blue,citecolor=blue]{hyperref}
\usepackage{amsmath}
\usepackage{graphicx}

\usepackage{dcolumn}
\usepackage{mathrsfs}
\usepackage{amssymb}
\usepackage{amsfonts,multirow}
\usepackage{bm,changes}
\usepackage{color}
\usepackage{ulem} 

\begin{document}
\draft
\title{Experimental demonstration of entanglement sudden death induced by
natural dissipation}
\author{Yan Wang}\thanks{These authors contribute equally to this work.}
\author{Hao-Long Zhang}\thanks{These authors contribute equally to this work.}
\author{Jia-Hao L\"{u}}
\author{Ken Chen}
\author{\\Wen Ning}\email{E-mail: ningw@fzu.edu.cn}
\affiliation{Fujian Key Laboratory of Quantum Information and Quantum Optics, College of Physics and Information Engineering, Fuzhou University, Fuzhou, Fujian 350108, China}
\author{Li-Hua Lin}
\author{Zhen-Biao Yang}\email{E-mail: zbyang@fzu.edu.cn}
\author{Shi-Biao Zheng}\email{E-mail: t96034@fzu.edu.cn}
\affiliation{Fujian Key Laboratory of Quantum Information and Quantum Optics, College of Physics and Information Engineering, Fuzhou University, Fuzhou, Fujian 350108, China}
\affiliation{Hefei National Laboratory, Hefei 230088, China}
\date{\today}

\begin{abstract}
Any quantum system inevitably interacts with its natural environment, which
can be modeled as a Markovian reservoir consisting of a continuum of
electromagnetic field modes. The quantum coherence of qubits in a
zero-temperature natural reservoir decays asymptotically, whereas the quantum entanglement of two qubits coupled to such reservoirs may disappear in a finite time. This phenomenon, referred to as entanglement sudden death
(ESD), has been simulated with artificially engineered dissipative channels,
but ESD induced by natural dissipative channels has not been confirmed.
We here present the first demonstration of natural-dissipation-induced ESD for two photonic
qubits, each stored in a leaky resonator of a superconducting circuit. The disentanglement dynamics of the two photonic qubits is monitored with two
ancilla superconducting qubits, which can be controllably coupled to the
corresponding leaky resonators. The techniques developed in our experiment pave the way for experimental exploration of entanglement dynamics in natural environments.
\end{abstract}

\maketitle

Quantum entanglement is a critical resource for implementation of various
quantum technologies, including quantum cryptography~\cite{PhysRevLett.67.661}, quantum
teleportation~\cite{PhysRevLett.70.1895}, and quantum sensing~\cite{RevModPhys.89.035002}. The main obstacle to
these technological applications is the environment-induced decoherence,
which turns a quantum-mechanically entangled state into a classically mixed
one. Understanding the disentanglement dynamics for qubits subjected to
decoherence is essential to mitigate decoherence effects. A
counterintuitive discovery is that the entanglement between two qubits can completely vanish in a finite time when each qubit is individually coupled
to a natural reservoir~\cite{PhysRevLett.93.140404}, where the qubit undergoes
irreversible spontaneous emission. This phenomenon, referred to as
entanglement sudden death (ESD), is somehow unexpected, as it takes an
infinitely long time for each qubit to completely decay from the excited
state to ground state in a Markovian vacuum reservoir consisting of a
continuum of electromagnetic modes with a flat spectrum. Following the
pioneering paper by Yu and Eberly~\cite{PhysRevLett.93.140404}, there have been a number of
theoretical works, devoted to various aspects of the entanglement dynamics during the
qubits-reservoirs interactions, including the ESD and entanglement sudden
rebirth (ESR) of two qubits each coupled to a cavity mode initially in the
vacuum state~\cite{Physics_B_2006,Physics_B_2007}, entanglement evolution of two qubits driven by coherent
fields~\cite{Yonac:08,Jarvis_2009}, disentanglement processes in non-Markovian reservoirs~\cite{PhysRevLett.99.160502,PhysRevA.79.042302},
and the ESR for reservoirs versus ESD for qubits~\cite{PhysRevLett.101.080503,XU20091906}.

On the experimental side, such processes have been demonstrated in optical
experiments, where Markovian~\cite{science.1139892,PhysRevA.78.022322} and non-Markovian ~\cite{PhysRevLett.104.100502} environments for
two initially entangled photonic qubits were simulated by classical optical
elements. ESD induced by dephasing noises has also been observed with two atomic
ensembles~\cite{PhysRevLett.99.180504}. On a solid-state platform, ESD and ESR were demonstrated with
a hybrid spin system controllably coupled to a spin bath~\cite{PhysRevB.98.064306}. However, none
of these experiments directly confirms the ESD induced by natural
reservoirs, as predicted in the original paper~\cite{PhysRevLett.93.140404}.

We here report an experimental demonstration of natural-dissipation-induced ESD with a
circuit quantum electrodynamics (QED) system, where two photonic modes
stored in separated microwave resonators are prepared in an entangled
state with two ancilla superconducting qubits. Each of these photonic qubits
is subjected to a dissipative channel, induced by a zero-temperature
Markovian reservoir, formed by infinitely many empty electromagnetic field modes. After a preset
disentanglement dynamics, the output density matrix of the two photonic
qubits is measured with the ancilla qubits. The results show that the
entanglement between the two photonic qubits can completely disappear within
a finite time under a certain condition, although the photons decay
asymptotically.

\begin{figure}[htbp]
  \centering
  \includegraphics{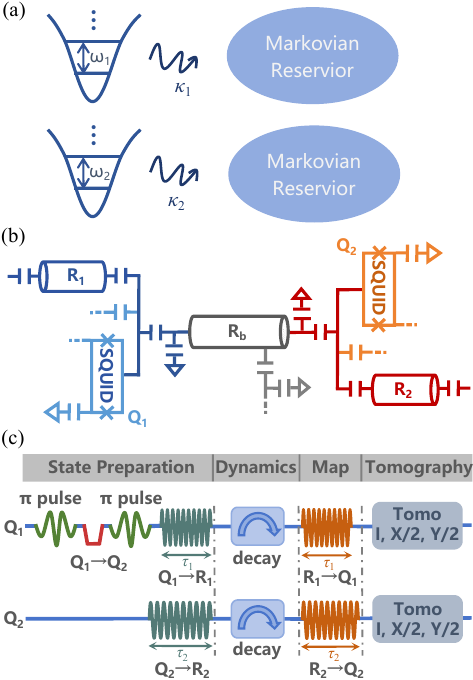}
  \caption{(color online). (a) Schematic of the naturally dissipative dynamics. Each of the two initially entangled photonic modes is individually coupled to a Markovian reservoir. The dissipation process is characterized by an irreversible photonic decay with the rate $\kappa _{j}$. (b) Experimental system. The disentanglement dynamics is tested with a circuit, where the readout resonators of two superconducting qubits Q$_{1}$ and Q$_{2}$, denoted as R$_{1}$ and R$_{2}$, act as the system photonic modes. Q$_{1}$ and Q$_{2}$, connected by a bus resonator (R$_{b}$), serve as ancilla qubits for preparing the initial entangled state of R$_{1}$ and R$_{2}$, and for measuring their output state. (c) Experimental pulse. The experiment involves three parts:
preparation of R$_{1}$-R$_{2}$ entangled state, disentanglement dynamics, and readout of R$_{1}$-R$_{2}$ state. R$_{1}$-R$_{2}$ entangled state is prepared with Q$_{1}$ and Q$_{2}$, which are entangled by subsequently transforming Q$_{1}$ from the ground state $\left\vert g\right\rangle _{1}$
to the excited state $\left\vert e\right\rangle _{1}$, partly transferring its excitation to Q$_{2}$ through the swapping coupling induced by virtual excitation of R$_{b}$, and flipping the state of Q$_{1}$. After these operations, Q$_{1}$ and Q$_{2}$ are evolved to the entangled state $\alpha
\left\vert g\right\rangle _{1}\left\vert g\right\rangle _{2}-\beta
\left\vert e\right\rangle _{1}\left\vert e\right\rangle _{2}$. Then the excitation of Q$_{j}$ is partly transferred to R$_{j}$ ($j=1,2$) through the sideband coupling mediated by parametric modulation. Due to the decoherence effects, the two readout resonators are prepared in the state of Eq.~\eqref{rhoq} by this transfer. After a preset disentanglement time $t$, the state of R$_{j}$ is mapped to Q$_{j}$, following which the Q$_{1}$-Q$_{2}$ density matrix is measured.
}\label{skemac}
\end{figure}

The system under consideration consists of two distinct photonic modes,
which are initially in an X-type mixed entangled state, whose density matrix
in the basis $\left\{ \left\vert 0\right\rangle _{1}\left\vert
0\right\rangle _{2},\left\vert 1\right\rangle _{1}\left\vert 0\right\rangle
_{2},\left\vert 0\right\rangle _{1}\left\vert 1\right\rangle _{2},\left\vert
1\right\rangle _{1}\left\vert 1\right\rangle _{2}\right\} $ is given by 
\begin{equation}\label{rhoq}
\rho =\left( 
\begin{tabular}{llll}
$v$ & 0 & 0 & $z$ \\ 
0 & $w$ & 0 & 0 \\ 
0 & 0 & $x$ & 0 \\ 
$z^{\ast }$ & 0 & 0 & $y$%
\end{tabular}%
\ \right) .
\end{equation}%
Here $\left\vert m\right\rangle _{1}\left\vert n\right\rangle _{2}$ denotes
the two-mode Fock state with $m$ $(n)$ photons in the first (second) mode. As
the each photonic mode is restricted in the subspace formed by zero- and
one-photon states, it can be considered a qubit. Each photonic mode is
subjected to a naturally dissipative channel induced by a Markovian
reservoir involving an ensemble of empty bosonic modes forming a
continuum with a flat spectrum. Under the Born-Markov approximation, the
freedom degrees of the reservoirs can be traced out. In the interaction
picture, the evolution of the density operator for the composite system
composed of the two photonic qubits is governed by the master equation~\cite{Scully_Zubairy_1997}%
\begin{equation}\label{drho}
\frac{d\rho }{dt}=\sum_{j=1}^{2}\frac{\kappa _{j}}{2}(2a_{j}\rho
a_{j}^{\dagger }-a_{j}^{\dagger }a_{j}\rho -\rho a_{j}^{\dagger }a_{j}),
\end{equation}%
where $a_{j}^{\dagger }$ and $a_{j}$ respectively denote the creation and
annihilation operators for the $j$th photonic mode with the decaying rate $%
\kappa _{j}$. The couplings between the photonic modes and the Markovian
 reservoirs are illustrated in Fig.~\ref{skemac}a. We note that such natural reservoirs
are the dominant decoherence source for quantum information processing with
logic qubits encoded in multi-photonic states~\cite{Nat.536.441,Nat.616.56}. After a disentanglement time $t$ the system's density operator evolves as

\begin{equation}
\rho ^{\prime }=\left( 
\begin{tabular}{llll}
$v^{\prime }$ & 0 & 0 & $z^{\prime }$ \\ 
0 & $w^{\prime }$ & 0 & 0 \\ 
0 & 0 & $x^{\prime }$ & 0 \\ 
$z^{\prime \ast }$ & 0 & 0 & $y^{\prime }$%
\end{tabular}%
\ \right) .
\end{equation}
where%
\begin{eqnarray}
v^{\prime } &=&v+\xi _{1}w+\xi _{2}x+\xi _{1}\xi _{2}y,  \nonumber \\
w^{^{\prime }} &=&\eta _{1}w+\eta _{1}\xi _{2}y,  \nonumber \\
x^{\prime } &=&\eta _{2}x+\xi _{1}\eta _{2}y,  \nonumber \\
y^{\prime } &=&\eta _{1}\eta _{2}y,  \nonumber \\
z^{\prime } &=&\sqrt{\eta _{1}\eta _{2}}z,
\end{eqnarray}%
with $\eta _{j}=e^{-\kappa _{j}t}$ and $\xi _{j}=1-\eta _{j}$ ($j=1,2$).

The concurrence~\cite{PhysRevLett.80.2245}, which quantifies quantum entanglement between these
two photonic qubits, is given by 
\begin{equation}
{\cal C}=2\max \{\left\vert z^{\prime }\right\vert -\sqrt{w^{\prime
}x^{\prime }},0\}.\text{ }
\end{equation}%
The two photonic qubits are initially in an entangled state when $\left\vert
z\right\vert >\sqrt{wx}$. During the natural decay, the entanglement would
disappear within a finite time when $\left\vert z\right\vert <\sqrt{
(w+y)(x+y)}$. The time for the occurrence of ESD is determined by
\begin{equation}\label{esd}
(w+y)(x+y)-wye^{-\kappa _{1}t}-xye^{-\kappa _{2}t}-\left\vert z\right\vert
^{2}+y^{2}e^{-(\kappa _{1}+\kappa _{2})t}=0 .
\end{equation}

The experiment is performed in a superconducting circuit consisting of five
frequency-tunable superconducting qubits, labeled as Q$_{j}$ ($j=1$ to $5$),
each connected to a microwave readout resonator (R$_{j}$), which holds an
electromagnetic field mode. The quantum states of different qubits can be
entangled with the assistance of a bus resonator (R$_{b}$), which has a
fixed frequency of $\omega_{b}/2\pi\approx 5.58$ GHz and a photonic decaying time T$_{b} \approx 13 $ $\rm \mu s$.
The field modes stored in R$_{1}$ and R$_{2}$ are used as photonic qubits to
demonstrate the ESD. The frequencies of these photonic modes are $\omega
_{1}^{r}/2\pi$ $\approx 6.65$ GHz and $\omega _{2}^{r}/2\pi$ $\approx 6.76$ GHz, and their decaying rates are $\kappa_{1}$ $\approx 1/240$ ${\rm ns}^{-1}$ and $\kappa _{2}$ $\approx 1/226$ ${\rm ns}^{-1}$, respectively. Q$_{1}$ and Q$_{2}$ serve as the
ancilla qubits for preparing the initial entangled state of these photonic
modes, and for reading out their state after the decoherence process. The
system is sketched in Fig.~\ref{skemac}b, where the unused three qubits and their
readout resonators are not shown. The on-resonance swapping coupling
strength between Q$_{1}$ (Q$_{2}$) and R$_{b}$ is $\lambda _{1}^{b}/2\pi$  ($\lambda _{2}^{b}/2\pi$) $\approx$ $21.4$ ($20.3$) MHz. The other three superconducting qubits stay at their
idle frequencies throughout the experiment, where they are highly detuned
and thus effectively decoupled from R$_{b}$. The energy relaxation time and
dephasing time of Q$_{1}$ (Q$_{2}$) are 11.2 (21.6) $\mu$s and 2.8 (0.9) $\mu$s, respectively. The detailed system parameters are shown in the Supplemental
Material~\cite{supplement}.

To establish the entanglement between these photonic modes, it is necessary to prepare Q$_{1}$ and Q$_{2}$ in an entangled state and then map this
entanglement to the R$_{1}$-R$_{2}$ system. The Q$_{1}$-Q$_{2}$ entangled state is synthesized by transforming Q$_{1}$ from the ground state $%
\left\vert g\right\rangle _{1}$ to the excited state $\left\vert e\right\rangle _{1}$ with a $\pi $ pulse at its idle frequency, and then tuning Q$_{1}$ and Q$_{2}$ to the frequency of  5.9 GHz, where a swapping coupling between Q$_{1}$ and Q$_{2}$ with a strength of $\xi/2\pi$ $\approx$ $1.4$ MHz is induced by virtual excitation of R$_{b}$~\cite{PhysRevLett.85.2392,PhysRevLett.119.180511}. After a preset interaction time $\tau 
$, the two superconducting qubits are evolved to the single-excitation entangled state $\alpha \left\vert e\right\rangle _{1}\left\vert
g\right\rangle _{2}-i\beta \left\vert g\right\rangle _{1}\left\vert
e\right\rangle _{2}$, where $\alpha =\cos (\xi \tau )$ and $\beta =\sin (\xi
\tau )$. Following this, both qubits are biased back to their idle
frequencies, where they are effectively decoupled from each other as the
detuning is much larger than the corresponding coupling strength. The subsequent $\pi $ pulse
on Q$_{1}$ transforms this entangled state to 
\begin{equation}
\alpha \left\vert g\right\rangle _{1}\left\vert g\right\rangle _{2}-i\beta
\left\vert e\right\rangle _{1}\left\vert e\right\rangle _{2}.
\end{equation}
The experimental pulse is shown in Fig.~\ref{skemac}c. As the coherence swapping rate
between Q$_{j}$ and R$_{b}$ is much larger than the corresponding
decoherence rates, the decoherence effects can be neglected during this
entanglement preparation. The fidelity of thus-prepared Q$_{1}$-Q$_{2}$
maximally entangled state is $98.5\%$.

The state of Q$_{j}$ ($j=1,2$) is partly transferred to R$_{j}$ through
their swapping coupling, mediated by an ac flux, which models the frequency
of Q$_{j}$ as $\omega _{j}^{\prime }=\omega _{j}+\varepsilon _{j}\cos (\nu
_{j}t)$. With the choice $\nu _{j}=\omega _{j}^{r}-\omega _{j}$, Q$_{j}$
is coupling to R$_{j}$ at the first-order upper sideband of the
corresponding parametric modulation~\cite{PhysRevApplied.10.054009,Nat.Phys.15.382,Nat.Commun.12.5924,PhysRevLett.131.113601,Nat.Commun.15.10293,PhysRevLett.131.260201}. The corresponding effective
interaction strength is given by $\chi _{j}=\lambda
_{j}J_{1}(\varepsilon _{j}/\nu _{j})$, where $J_{1}(x)$ is the first-order
Bessel function of the first kind, and $\lambda _{1}$ $(\lambda _{2})$ $\equiv 2\pi \times$ 41.9 (40.0) MHz
is the on-resonance coupling between Q$_{1}$ (Q$_{2}$) and R$_{1}$ (R$_{2}$%
). In our experiment, the two sideband interaction strengths are $\chi _{1} \equiv 2\pi \times 4.41$  MHz and $\chi _{2} \equiv 2\pi\times 3.18$ MHz, respectively. After the interaction time $\tau
_{j}= (\pi-\arctan (4\Omega _{j}/\chi _{j}))/\Omega_j$ with $\Omega _{j}=\sqrt{\chi
_{j}^{2}-\kappa _{j}^{2}/16}$, the excitation of Q$_{j}$ is partly
transfered to R$_{j}$, and partly leaked to the environment. The coherence
times of the ancilla qubits are much longer than the corresponding state
transfer times, and thus the decoherence effects of these qubits can be
neglected. Consequently, the two photonic modes are prepared in the mixed
state of the form of Eq.~\eqref{rhoq}, where%
\begin{eqnarray}
v &=&\alpha ^{2}+(1-p_{1})(1-p_{2})\beta ^{2}  \nonumber, \\
w &=&p_{1}(1-p_{2})\beta ^{2},  \nonumber \\
x &=&(1-p_{1})p_{2}\beta ^{2},  \nonumber \\
y &=&p_{1}p_{2}\beta ^{2},  \nonumber \\
z &=&-i\sqrt{p_{1}p_{2}}\alpha \beta ,
\end{eqnarray}%
with $p_{j}=[\chi _{j}e^{-\kappa _{j}\tau _{j}/4}\sin (\Omega _{j}\tau
_{j})/\Omega _{j}]^{2}$. Such a mixed state has a nonvanishing entanglement
when $\beta \neq 0$ and $\sqrt{(1-p_{1})(1-p_{2})}\beta <\alpha $. Upon this
entanglement transfer, the two ancilla qubits are left in the ground state. Theoretically, ESD can occur for $\left\vert \beta\right\vert > 1/\sqrt{2}$.

After this state transfer, the parametric modulations are switched off so
that Q$_{j}$ is decoupled from R$_{j}$. Then each of the two photonic modes
evolves under natural dissipation, with the system dynamics being described
by Eq.~\eqref{drho}. After a preset disentanglement time $t$, the output $R_{1}$-$%
R_{2}$ density matrix is measured by re-applying the parametric modulations
to the two ancilla qubits, mediating the corresponding sideband interactions
between them and the readout resonators. 
After the interaction time $\tau
_{j}$, 
\begin{figure}[htbp]
  \centering
    \includegraphics{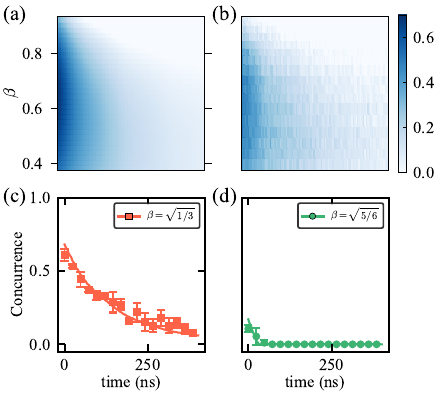}
  \caption{(color online). (a) Simulated concurrence versus $\beta $ and $t$. (b) Measured concurrence versus $\beta $ and $t$. (c), (d) Evolutions of the measured concurrence for $\beta = \sqrt{1/3}$ (c) and $\sqrt{5/6}$
(d). The solid lines are the results calculated with the master equation.}\label{concurrence_t}
\end{figure}
\begin{figure}[htbp]
  \centering
  \includegraphics{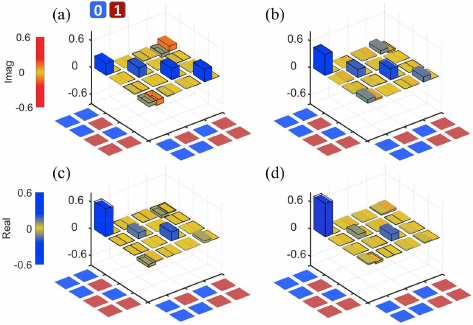}
  \caption{(color online). (a)-(d)   $R_{1}$-$R_{2}$ density matrices measured for disentanglement times $t=$ 100, 200, 300, and 400 $\mathrm{ns}$.  The results are obtained for $\beta =\sqrt{5/6}$. Each matrix element is characterized by two color bars, one for the real part and the other for the imaginary part. The black wire frames denote the matrix elements of the ideal output states.}\label{visually}
\end{figure}
the state of $R_{j}$ is partly transferred to $Q_{j}$. Following this
state mapping, the parametric modulations are switched off, and the $Q_{1}$-$%
Q_{2}$ density matrix is reconstructed by quantum state tomography. We can
infer the joint R$_{1}$-R$_{2}$ state just before the state mapping from the
measured output Q$_{1}$-Q$_{2}$ density matrix, as detailed in the
Supplemental Material. Fig.~\ref{concurrence_t}a presents the concurrence calculated with thus-obtained R$_{1}$-R$_{2}$ density matrix, versus $\beta $ and $t$. As expected, the concurrence exhibits an asymptotic decaying process when $\beta $ is small, but suddenly drops to zero when $\beta $ is sufficiently large. Furthermore, the ESD occurs at an earlier time when $\beta $ is increased. The experimental result well agrees with the nemurical simulation, shown in Fig.~\ref{concurrence_t}b. To more clearly show the difference between
these two disentanglement process, in Fig.~\ref{concurrence_t}c and \ref{concurrence_t}d we present the
concurrences evolutions for $\beta =\sqrt{1/3}$ and $\sqrt{5/6}$, respectively.

To gain a deeper understanding of the physics underlying the ESD, in Fig.~\ref{visually}a-d we display the output density 
\begin{figure}[htbp]
  \centering
  \includegraphics{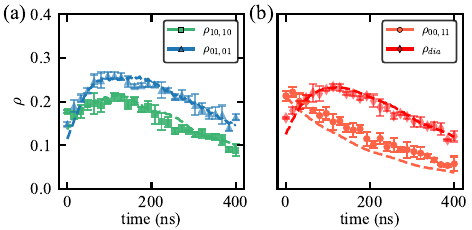}
  \caption{(color online). (a) Observed evolutions of the diagonal elements $\rho _{10,10}$ and $\rho _{01,01}$. (b) Observed evolutions of $\left\vert \rho _{00,11}\right\vert $ and $\rho_{dia}=\sqrt{\rho _{10,10}\rho _{01,01}}$. The results are obtained for $\beta =\sqrt{5/6}$. The dashed lines are the results calculated with the master equation.}\label{rhot}
\end{figure}
\begin{figure}[htbp]
  \centering
  \includegraphics{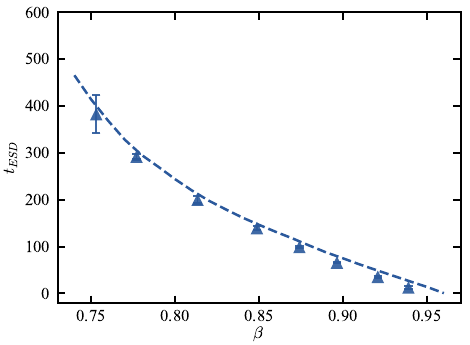}
  \caption{(color online). ESD time ($t_{ESD}$) versus $\beta$. For each value of $\beta $, $t_{ESD}$ is equal to the transverse coordinate of the cross point between the observed $\left\vert\rho _{00,11}\right\vert $ and $\rho _{10,10}^{01,01}$ as functions of $t$. The dashed lines are the results calculated with the master equation.}\label{tesd}
\end{figure}
matrices of the two photonic qubits for $\beta =\sqrt{5/6}$, measured for different disentanglement times $t=$ 100, 200, 300, and 400 $\mathrm{ns}$.
As the entanglement is irrelevant to the phases of the off-diagonal elements 
$\rho _{00,11}$ and $\rho _{11,00}$, we here only display their magnitude.
The disentanglement dynamics depends upon the interplay between these
off-diagonal elements and diagonal elements, $\rho _{10,10}$ and $\rho
_{01,01}$. Fig.~\ref{rhot}a shows the observed temporal behaviors of $\rho _{10,10}$
and $\rho _{01,01}$. As expected, these two diagonal elements exhibit
non-monotonous behaviors. Fig.~\ref{rhot}b presents the observed evolution of $%
\left\vert \rho _{00,11}\right\vert $, together with the evolution of the
geometric mean value of measured diagonal elements $\rho _{10,10}$ and $\rho _{01,01}$, 
\begin{equation}
\rho _{dia}=\sqrt{\rho _{10,10}\rho _{01,01}}.
\end{equation}

For $\beta >1/\sqrt{2}$, there exists a cross point between the functions $%
\left\vert \rho _{00,11}(t)\right\vert $ and $\rho _{dia}(t)$,
whose transverse coordinate corresponds to the time for the occurrence
of ESD. The ESD times for different values of $\beta $ that are larger $1/
\sqrt{2}$ can be inferred from the observed evolutions of the output $R_{1}$-%
$R_{2}$ matrices measured for different decaying times. Fig.~\ref{tesd} presents
thus-obtained ESD time versus $\beta $, which is in well agreement with the
result calculated with Eq.~\eqref{esd} (dashed line).

In conclusion, we have presented an experimental demonstration of
ESD for two photonic qubits, each subjected to an
irreversible dissipative channel induced by a natural reservoir. The
experiment is performed with a circuit QED system, where the photonic modes
are stored in two leaky microwave resonators. Two superconducting qubits connected
by a bus resonator serve as ancillae for preparing and reading out the state
of the photonic modes. The experimental results well agree with theoretical predictions, confirming that
ESD can occur for a wide range of initial entangled state. The techniques developed in the experiment can be used to investigate the dynamics for complex entangled states of photonic modes in natural reservoirs.

This work was supported by the National Natural Science Foundation of China (Grants No. 12474356, No. 12475015, No. 12274080, and No. 12505016), Quantum Science and Technology-National Science and Technology Major Project (Grant No. 2021ZD0300200), and the Natural Science Foundation of Fujian Province (Grant No. 2025J01465). 


\nocite{y2007,Nat.Commun.8.1061,PhysRevLett.123.060502,PhysRevB.87.220505,PhysRevLett.131.260201}
\bibliography{reference}
\end{document}


\title{Supplementary Material for ``Experimental demonstration of entanglement sudden death induced by
natural dissipation"}
\author{Yan Wang}\thanks{These authors contribute equally to this work.}
\author{Hao-Long Zhang}\thanks{These authors contribute equally to this work.}
\author{Jia-Hao L\"{u}}
\author{Ken Chen}
\author{\\Wen Ning}\email{E-mail: ningw@fzu.edu.cn}
\affiliation{Fujian Key Laboratory of Quantum Information and Quantum Optics, College of Physics and Information Engineering, Fuzhou University, Fuzhou, Fujian 350108, China}
\author{Li-Hua Lin}
\author{Zhen-Biao Yang}\email{E-mail: zbyang@fzu.edu.cn}
\author{Shi-Biao Zheng}\email{E-mail: t96034@fzu.edu.cn}
\affiliation{Fujian Key Laboratory of Quantum Information and Quantum Optics, College of Physics and Information Engineering, Fuzhou University, Fuzhou, Fujian 350108, China}
\affiliation{Hefei National Laboratory, Hefei 230088, China}
\date{\today}

\maketitle
\tableofcontents

\section{Calculation of Pairwise Concurrences}
We prepare the initial qubit state and then swap such a state into the resonators following the procedure outlined in the main text. During this transfer process, part of the excitation of $Q_j$ is coherently transferred to $R_j$, while the remaining fraction is dissipated into the environment. The system dynamics is governed by the following
\begin{equation}
\frac{d\rho }{dt}= \mathscr{L}[\rho] =\sum_{j=1}^{2} -\frac{i}{\hbar}[H_j,\rho]+  \frac{\kappa _{j}}{2}(2a_{j}\rho
a_{j}^{\dagger }-a_{j}^{\dagger }a_{j}\rho -\rho a_{j}^{\dagger }a_{j}),
\end{equation}
with $H_j=\lambda_j(\sigma_j^-a_j^\dagger + \sigma_j^+a_j)$, where $\sigma^{\dagger}_{j}=|e\rangle_{j}\langle g|$, $\sigma^{-}_{j}=|g\rangle_j\langle e|$, the operator $a_{j}^{\dagger}$ is the creation operator for the resonator coupled to $Q_j$ ($j=1,2$), and $\lambda_{j}$ represents the coupling strength between $Q_j$ and $R_j$ .

Since the two subsystems are decoupled, we independently solve their respective eigenvalue equations, $\mathscr{L}\tilde\rho_i = E_i\tilde\rho_i$, to obtain the corresponding eigenvalues and eigenstates. The initial state of the system can then be expanded in this eigenbasis as
\begin{equation}
\tilde\rho(0) = \sum_{m,n}C_{m,n}\tilde\rho_m \otimes \tilde\rho_n.
\end{equation}

After an evolution time of $t$, where $t$ is the swap duration needed to transfer the information of $Q_j$ to $R_j$, the system's density matrix becomes
\begin{equation}
\tilde\rho(t) = \sum_{m,n}C_{m,n}e^{(E_m + E_n)t}\tilde\rho_m \otimes \tilde\rho_n.
\end{equation}

Consequently, by tracing out the qubit degrees of freedom, we obtain the desired mixed-state reduced density matrix for the resonators, given in Eq.~(1) of the main text.

To quantify the entanglement, we consider a two-qubit mixed state in the standard \textquotedblleft X\textquotedblright\ form~\cite{y2007}

\begin{equation}
\rho _{A,B}=\left(
\begin{tabular}{llll}
$a$ & 0 & 0 & $z$ \\
0 & $b$ & 0 & 0 \\
0 & 0 & $c$ & 0 \\
$z^{\ast }$ & 0 & 0 & $d$%
\end{tabular}%
\ \right) .
\end{equation}
For this class of states, the concurrence is
\begin{equation}\label{conc}
{\cal C}_{A,B}=2\max \{\left\vert z\right\vert -bc\text{, }0\}.\text{}
\end{equation}
The reduced density matrix in Eq.~(3) of the main text confirms exactly to this X form, where the parameters $ a, b, c,d $ and $ z $ are given by Eq.~(4). Substituting these expressions into Eq.~\eqref{conc}, we directly obtain $ {\cal C}_{r_{1},r_{2}} $ , as shown in Eq.~(5).

\section{Experimental Setup and Device Information}

\begin{figure}[htbp]
  \centering
    \includegraphics[width=0.75\linewidth]{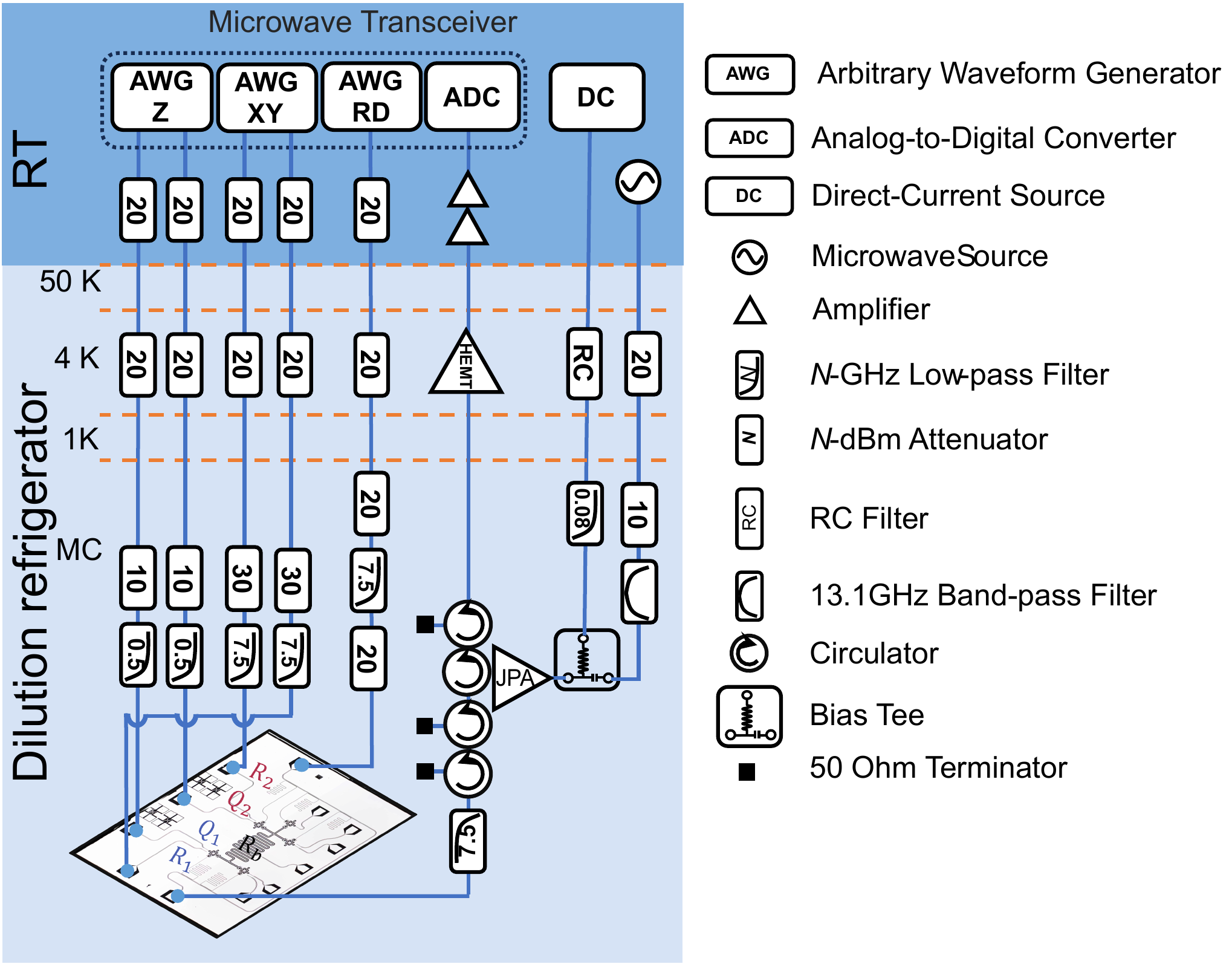}
    \caption{A schematic diagram of the experimental system and wiring information.}
    \label{deviceinfo}
\end{figure}

Figure~\ref{deviceinfo} shows a complete schematic of the wiring and electronics from room temperature (RT) down to the low-temperature stages inside the dilution refrigerator. The chip is mounted in a BlueFors dilution refrigerator, with the mixing chamber (MC) stage maintained at a base  temperature below 20 mK.

The experimental sample consists of a bus resonator and five Xmon qubits~\cite{Nat.Commun.8.1061,PhysRevLett.123.060502}. The bus resonator has a fixed bare frequency $\omega_\text{p}/2\pi = 5.581$ GHz and a photonic decay time $T_{1,\text{p}} = 12.9$ $\mu$s. Each qubit is coupled to a dedicated readout resonator, with frequencies in the range 6.6--6.9 GHz, for state readout. Moreover, each qubit is equipped with two control lines: an XY line for resonant driving and a Z line for frequency tuning. The Z control, in conjunction with time-dependent drives, enables the qubit to couple to both its readout resonator and the bus resonator.

Qubit control and reaout pulses are generated by an integrated arbitrary waveform generator (AWG). The AWG provides XY channels, which output pulses in the 0.1-9.8 GHz range for nanosecond-scale qubit flipping, and Z channels, which deliver dc-1.8 GHz pulses for fast frequency tuning. For readout, the AWG generate simultaneous multitone pulses across the 0.1-9.8 GHz band to address all qubit readout resonators. The output signal is sequentially amplified by an impedance-transformed Josephson parametric amplifier (JPA), a high electron mobility transistor (HEMT), and room temperature amplifiers, then digitized and demodulated by the AWG’s analog-to-digital converters (ADCs). All AWG channels are controlled by a field-programmable gate array with nanosecond latency. The JPA is pumped by a microwave source at approximately 13.5 GHz and operated with a DC bias. To reduce noise and prevent spurious reflections, circulators, attenuators, and filters are inserted into the signal lines at the MC and 4K stages of the dilution refringerator, as well as at RT layer.

We summarize the qubit and resonator parameters in Table~\ref{table1}. Two of the five qubits, $Q_1$ and $Q_2$, are initially prepared in a specific entangled state via their dispersive coupling to the bus resonator $R_\text{b}$. This entanglement is then resonantly mapped to the readout resonators $R_1$ and $R_2$ using modulated drives on the Z control lines of both qubits, while continuously undergoing dissipation from single-photon loss in $R_1$ and $R_2$. The ensuing dynamics is monitored in real time via joint quantum state tomography on $Q_1$ and $Q_2$, revealing the onset of entanglement sudden death. Details of superconducting devices and
control techniques
follow those reported in Ref.~\cite{PhysRevB.87.220505,PhysRevLett.131.260201}.

\begin{table}[h]
    \centering
    \begin{tabular*}{\textwidth}{@{\extracolsep{\fill}} c c c}
        \toprule
        Parameters & $Q_1$ & $Q_2$ \\
        \midrule
        Qubit idle frequency, $\omega_{id,j}/2\pi$ & 5.90 GHz & 5.85 GHz \\
        Coupling strength to the bus resonator $R_{b}$, $g_{j,b}/2\pi$ & 21.4 MHz & 20.3 MHz\\
        Coupling strength to the readout resonator $R_{j,r}$, $g_{j,r}/2\pi$ & 41 MHz & 40 MHz \\
        Energy relaxation time, $T_{1,j}$ & 11.2 $\mu$s & 21.6 $\mu$s \\
        Ramsey dephasing time, $T_{2,j}^*$ & 2.8 $\mu$s & 0.9 $\mu$s \\
        Dephasing time with spin echo, $T_{2,j}^{SE}$ & 13.6 $\mu$s & 3.5 $\mu$s \\
        Frequency of readout resonator, $\omega_{r,j}/2\pi$ & 6.65 GHz & 6.76 GHz \\
        Leakage rate of readout resonator, $\kappa_{j}$ & $1/240$ $\rm ns^{-1}$ & $1/226$ $\rm ns^{-1}$ \\
        Measure fidelity for the qubit, $F_{0,j}$ ($F_{1,j}$) & $0.89$ ($0.84$) & $0.90$ ($0.82$) \\
        \bottomrule
    \end{tabular*}
    \caption{Characterized parameters of the superconducting circuit in experiment (those for the two used qubits, $Q_1$ and $Q_2$, are shown). $\omega_{id,j}/2\pi$ is the qubit idle frequency, where the qubits stay before and after their interaction with the bus resonator or the readout resonator; $\omega_{id,j}/2\pi$ is also the frequency position where the single-qubit manipulation and the qubit state measurement are performed. $g_{j,b}$ ($g_{j,r}$) denotes the constant coupling strength between $Q_j$ and the bus resonator (its readout resonator), $R_b$ ($R_{j,r}$), estimated by means of resonant $Q_j-R_b$ (dispersive $Q_j-R_{j,r}$) interaction. $T_{1,j}$ and $T_{2,j}^*$ are the energy relaxation time and Ramsey dephasing time (Gaussian dey) of $Q_j$, respectively, both of which are measured at the qubit idle frequency. $T_{2,j}^{SE}$ is the dephasing time (Gaussian decay) with spin echo for $Q_j$. $\omega_{r,j}/2\pi$ and $\kappa_{f,j}$ represent the fixed frequency and leakage rate of the readout resonator, $R_{j,r}$, respectively. $F_{0,j}$ ($F_{1,j}$) is the measure fidelity for $Q_j$, which characterizes the probability of correctly measuring $Q_j$ in $|0\rangle$ ($|1\rangle$) when it is prepared in $|0\rangle$ ($|1\rangle$) state.}
    \label{table1}
\end{table}

\section{Implementation of Qubit-Readout Resonator Couplings}
In our experiment, energy exchange between each qubit $Q_j$ and its readout resonator $R_j$ is mediated by applying an ac flux to the qubit, with the qubit frequency modulated as~\cite{PhysRevB.87.220505,PhysRevLett.131.260201}
\begin{equation}
    \omega_{e,j}(t)\approx\omega_{0,j}+\varepsilon_j\cos(\nu_j t),
    \label{S9}
\end{equation}
where $j=1,2$, $\omega_{0,j}$  denotes the average frequency of the $j$-th qubit and $\varepsilon_j$ and $\nu_j$ are the modulation amplitude and frequency for qubit $Q_j$, respectively.
Under this parametric modulation, the coherent dynamics of the combined Q-R system is governed by the Hamiltonian (setting $\hbar$ = 1)
\begin{equation}
   H=\sum_{j=1,2}\omega_{e,j}(t)\sigma_{z.j}/2+\omega_{r,j}a^\dagger_j a_j+\lambda_j(a^\dagger_j\sigma_j^-+a_j\sigma_j^\dagger),
\end{equation}
where $\sigma_{z,j}=|e\rangle_{j}\langle e|-|g\rangle_{j}\langle g|$, $\sigma^{\dagger}_{j}=|e\rangle_{j}\langle g|$, $\sigma^{-}_{j}=|g\rangle_j\langle e|$, $\omega_{r,j}$ is the center frequency of the readout resonant and $\lambda_{j}$ represents the coupling strength between qubit $Q_j$ and readout resonator $R_j$, $j=1,2$. In the rotating frame, the interaction Hamiltonian can be transformed into

\begin{equation}
    H_I=\sum_{j=1,2} \lambda_je^{i\Delta_jt-i\mu_j\sin(\nu_j t)}a^\dagger_j\sigma_j^-+H.c.,\label{SHI}
\end{equation}
where $\Delta_j = \omega_{r,j} - \omega_{0,j}$ , $\mu_j = \varepsilon_j/\nu_j$, and $ H.c.$ is the Hermitian conjugate. With the application of the Jacobi-Anger expansion $e^{i\mu\sin\theta}=\sum_{-\infty}^{\infty}J_n(\mu)e^{in\theta}$, where $J_n(\mu)$ denotes the $n$-th Bessel function of the first kind, Eq.~(\ref{SHI}) can be rewritten as
\begin{equation}
   H_I=\sum_{j=1,2}\lambda_j\left[\sum_{n=-\infty}^\infty J_n(\mu)e^{-i(n\nu_j-\Delta_j)t}a^\dagger\sigma^-_j+H.c.\right].
   \label{S12}
\end{equation}

To achieve the first-order sideband modulation of the $Q$-$R$ coupling, we set $\Delta_j = \nu_j$ and $\lambda_j \ll \nu_j$. In this case, terms with $n\neq1$, which oscillate rapidly, can be neglected under the rotating-wave approximation. As a result, the Hamiltonian Eq.~(\ref{S12}) reduces to $H_I^{'}=\lambda_jJ_1(\mu)a^\dagger\sigma^-_j + H.c.$, where the coupling strength between $Q_j$ and $R_j$ is given by $g_j = \lambda_j J_1(\mu)$.

\begin{figure}[htbp]
  \centering
  \includegraphics{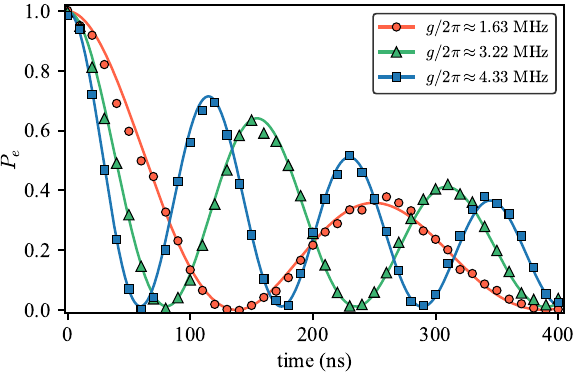}
  \caption{(color online). The time evolution of the excited state population under longitudinal modulations with different parameters. The dots represent the experimental results and the solid lines are the fitting results.
}\label{S2}
\end{figure}

To realize the $Q$-$R$ coupling, we first prepare the qubit in the excited state $|e\rangle$ and apply longitudinal modulation as described in Eq.~(\ref{S9}). When $\varepsilon$ and $\nu$ are properly tuned, a complete exchange between $Q$ and $R$ is observed, enabling a swap operation $|e,0\rangle\rightarrow|g,1\rangle$ under this modulation scheme.
The coupling strength $g_j$ for a given set of parameters $\nu_j$, $\varepsilon_j$ and $\omega_j$ can be extracted through numerical fitting. To validate the effectiveness of this modulation approach, Fig.~\ref{S2} displays the vacuum Rabi oscillations in qubit $Q_1$, driven by sideband couplings under varying modulation amplitudes. 

\section{State Preparation and Readout of Photonic Modes}

As mentioned in the main text, directly preparing and reading out quantum states in the readout resonator is difficult. Therefore, we first prepare the target entanglement state on the qubits, and then transfer the state to the corresponding readout resonator via sideband coupling. Subsequently, the state is mapped back to the qubits, and the output density matrix is reconstructed via joint quantum state tomography. The first-order sideband coupling strengths of the two qubits with their respective readout resonators are $\lambda_1 \simeq 2\pi \times 4.41 $ MHz and $\lambda_2 \simeq 2\pi \times 3.18$ MHz, and the corresponding interaction times are $\tau_1 = 58.8\ n$s  and  $\tau_2 = 82.4\ n$s, respectively. The same parameters are employed for the mapping process.
To precisely determine the $Q_j-R_j$ interaction time, we individually prepare $Q_j$ in the excited state $|e\rangle _j$, measure their vacuum Rabi oscillation signals under the first-order sideband coupling with the $R_j$, and take the time at which the qubit population first completely resides in the ground state $|g\rangle_j$ as the interaction time $\tau_j$ between the qubit and the readout resonator.

\begin{figure}[htbp]
  \centering
  \includegraphics{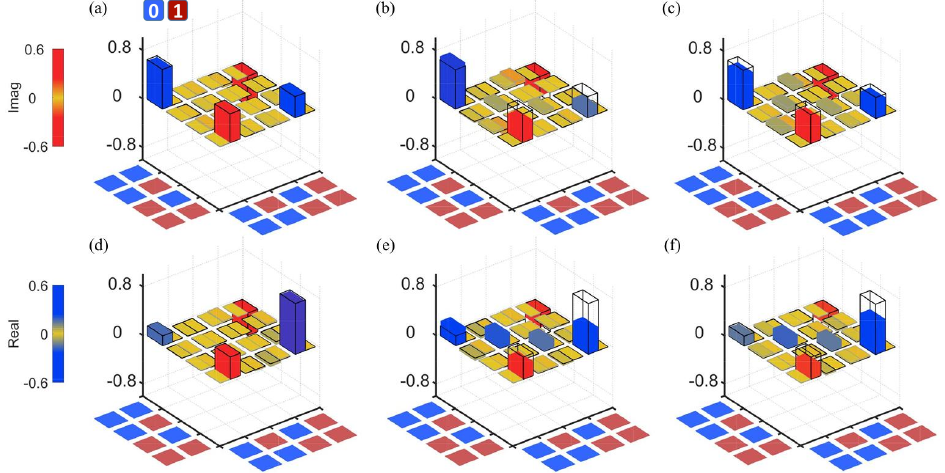}
  \caption{(color online). Reconstructed density matrices for $\beta = \sqrt{1/3}$ and $\sqrt{5/6}$. (a) (d) The initially prepared states of $Q_1$-$Q_2$. (b) (e) The output states of $Q_1$-$Q_2$ after the sequential state transfer to $R_1$-$R_2$ and back. (c) (f) The inferred density matrices of $R_1$-$R_2$ just before the readout mapping, reconstructed from the measured output states of $Q_1$-$Q_2$. Each matrix element is characterized by two color bars, one for the real part and the other for the imaginary part. The black wire frames denote the matrix elements of the ideal initially prepared states.}\label{S3}
\end{figure}

We present the density matrices of $Q_1$-$Q_2$, as well as those after the sequential state transfer from $Q_1$-$Q_2$ to $R_1$-$R_2$ and back to $Q_1$-$Q_2$ in Fig.~\ref{S3}a-b (d-e), for $\beta=\sqrt{1/3}$ ($\sqrt{5/6}$), respectively.
The fidelity of the initial two-qubit state and the mapped two-qubit density matrix are 0.974 (0.961) and 0.821 (0.500), with the corresponding concurrences are 0.924 (0.726) and 0.513 (0.118), respectively.
These results demonstrate that the dissipation of the readout resonator induces non-negligible decoherence of the quantum state during the state mapping process.
To improve  agreement between the numerical simulation results and the experimental observations in the main text,
we infer the joint $R_{1}$-$R_{2}$ state just before the state mapping from the
measured output $Q_1$-$Q_2$ density matrix~\cite{PhysRevLett.131.260201}, shown in Fig.~\ref{S3}c (f).
The corresponding fidelity and concurrence are 0.827 (0.578) and 0.610 (0.106), respectively. Compared with the mapped two-qubit density matrix, both the fidelity and the concurrence exhibit a moderate improvement, indicating that this approach corrects for dissipative effects introduced during the mapping process.


For clarity, we take a pure state as an example to show how to infer the state of $R_1$ from the output state $Q_1$. The state of $R_1$ before the mapping can be expressed as
\begin{equation}
    |\psi_0\rangle = \alpha |0\rangle + \beta |1\rangle,
 \end{equation}
with $|\alpha^2| + |\beta|^2 =1$. After the state mapping, the $Q_1$ output state is,
\begin{equation}
    |\psi_1\rangle = (1/\sqrt{|\alpha|^2 + k_1^2|\beta|^2})(\alpha|g\rangle + k_1 \beta|e\rangle) ,
\end{equation}
where $k_1 = e^{-\kappa_1 \tau_1/2}$. This implies that the original $R_1$ output state $|\psi_0\rangle$ can be inferred from $|\psi_1\rangle$ by multiplying the coefficient of the component $|e\rangle$ by $1/k_1$ and then renormalizing the resulting state. The result associated with the no-jump trajectory is obtained by projecting the density matrix to the single-excitation subspace $\{|g\rangle, |e\rangle\}$, which can be expressed as

\begin{equation}
\rho=\left(
\begin{tabular}{ll}
$\rho_{00}$ & $\rho_{01}$ \\
$\rho_{10}$ & $\rho_{11}$ \\
\end{tabular}%
\ \right).
\end{equation}
The elements of $R_1$ density matrix within $\{|0\rangle, |1\rangle\}$ right before the state mapping are related to those of $\rho_1$ by
\begin{equation}
    \rho_1 = \frac{1}{\rho_{11} + \rho_{22}/k^2}
    \begin{pmatrix}
        \rho_{11} & \rho_{12}/k \\
        \rho_{21}/k & \rho_{22}/k^2
    \end{pmatrix}.
\end{equation}

For the joint $R_1$-$R_2$ mixed  state, we project it onto the basis $\left\{ \left\vert 0_{1}0_{2}\right\rangle,\left\vert 1_{1}0_{2}\right\rangle,\left\vert 0_{1}1_{2}\right\rangle,\left\vert
1_{1}1_{2}\right\rangle\right\} $. By applying a similar procedure and accounting for the dissipation of both resonators, we can deduce the results we need.  All the data presented in the main text are processed using this approach.

\bibliography{reference}